\documentclass[a4paper,fleqn]{cas-dc}

\usepackage[numbers,sort&compress]{natbib}

\def\tsc#1{\csdef{#1}{\textsc{\lowercase{#1}}\xspace}}
\tsc{WGM}
\tsc{QE}

\usepackage{amsmath,amssymb,bm}
\usepackage{graphicx}
\usepackage[dvipsnames]{xcolor}
\usepackage{hyperref}
\usepackage{url}
\usepackage{ragged2e}
\usepackage{array}
\usepackage{comment}

\definecolor{DARKBLUE}{HTML}{00008b}
\hypersetup{
colorlinks=true,
urlcolor=DARKBLUE,
anchorcolor=DARKBLUE,
citecolor=DARKBLUE,
filecolor=DARKBLUE,
linkcolor=DARKBLUE,
menucolor=DARKBLUE,
linktocpage=true
}

\newcommand{\rd}{r_{\mathrm d}}
\newcommand{\rdf}{r^{\mathrm{fid}}_{\mathrm d}}
\newcommand{\dm}{D_{\mathrm M}}

\newcommand{\dmf}{D^{\mathrm{fid}}_{\mathrm M}}
\renewcommand{\dh}{D_{\mathrm H}}
\newcommand{\dhf}{D^{\mathrm{fid}}_{\mathrm H}}

\newcommand{\nn}{\nonumber}

\begin{document}
\let\WriteBookmarks\relax
\def\floatpagepagefraction{1}
\def\textpagefraction{.001}

\shorttitle{The cosmic tetrarchy}    

\shortauthors{M. Martinelli, D. Sapone}  

\title [mode = title]{The cosmic tetrarchy: four estimators breaking the assumption degeneracy in cosmological distance tensions}

\author[1,2]{Matteo Martinelli}[orcid=0000-0002-6943-7732]\ead{matteo.martinelli@inaf.it}
\credit{Conceptualization, Methodology, Software, Formal analysis, Investigation, Validation, Visualization, Writing – original draft, Writing – review and editing}

\affiliation[1]{organization={INAF - Osservatorio Astronomico di Roma},
            addressline={via Frascati 33}, 
            city={Monteporzio Catone
(Roma)},
            postcode={00040}, 
            country={Italy}}

\affiliation[2]{organization={INFN - Sezione di Roma},
            addressline={Piazzale Aldo Moro, 2 - c/o Dipartimento di Fisica, Edificio G. Marconi}, 
            city={Roma},
            postcode={00185}, 
            country={Italy}}

\author[3]{Domenico Sapone}[orcid=0000-0001-7089-4503]
\cormark[1]
\cortext[1]{Corresponding author}
\ead{domenico.sapone@uchile.cl}

\credit{Conceptualization, Methodology, Software, Formal analysis, Investigation, Visualization, Writing – original draft, Writing – review and editing}

\affiliation[3]{organization={Departamento de F\'isica, FCFM, Universidad de Chile},
            addressline={Blanco Encalada 2008}, 
            city={Santiago},
            country={Chile}}

\begin{abstract}
The origin of cosmological distance tensions remains a central open question in precision cosmology, complicated by the fact that most consistency tests between datasets cannot isolate which physical assumption is responsible for an observed discrepancy. We address this by reformulating the standard cosmological framework as a single null test: the requirement that the dimensionless sound-horizon ratio $\rd/\rdf$ be one redshift-independent number. 
We show that this test admits four complementary measurements, obtained by combining Baryon Acoustic Oscillation (BAO) data with either Type Ia supernovae (SNIa) or cosmic chronometers (CC), in either the transverse or the radial direction. The four channels rely on distinct subsets of physical assumptions --- distance-ladder calibration, the distance duality relation, spatial flatness, and the standard-ruler picture --- and one of them, the radial CC-anchored channel, requires none and serves as the natural reference of the framework. 
The pattern of agreement or disagreement among the four therefore localises the assumption responsible for any observed tension. We refer to this fourfold decomposition as the \emph{cosmic tetrarchy} and evaluate it on DESI DR2 BAO data combined with Pantheon+ and cosmic chronometers, using both a binned analysis with full analytic covariance propagation and a non-parametric Gaussian Process reconstruction.
We find that current data are compatible with a single, redshift-independent sound-horizon scale; when comparing the different estimators, we find hints of discrepancies between those based on SNIa and those relying on CC, although with a lack of statistical significance, which might hint to a manifestation of the Hubble tension or to the breaking of assumptions such as the distance duality relation.
\end{abstract}





\maketitle

\section{Introduction}\label{sec:intro}

The current era of precision cosmology is characterised by an unprecedented wealth of high-quality observations, enabling increasingly stringent tests of the standard cosmological model. Measurements of the cosmic microwave background~\cite{Planck:2018vyg}, large-scale structure~\cite{DESI:2025zgx, BOSS:2012dmf}, and distance indicators~\cite{Jimenez:2001gg} have reached percent-level precision, allowing detailed reconstruction of the expansion history of the Universe (see the reviews \cite{DiValentino:2021izs, CosmoVerseNetwork:2025alb} and references therein). At the same time, this progress has revealed a number of persistent tensions between independent probes, most notably in the determination of the Hubble constant and in the calibration of cosmological distances \cite{Riess:2021jrx}.

A key limitation in interpreting these tensions lies in their dependence on underlying model assumptions. Most cosmological inferences rely on fitting parametric models (typically within the $\Lambda$CDM framework) to heterogeneous datasets. While this approach is powerful, it can obscure the origin of discrepancies, as tensions may arise from a combination of physical effects, hidden systematics, or implicit assumptions such as spatial flatness, photon conservation, or the validity of the distance duality relation (DDR). Disentangling these possibilities requires complementary approaches that minimise model dependence and instead focus on direct consistency relations among observables.

Several model-independent consistency tests have been proposed in the literature to probe specific assumptions underlying the standard cosmological framework. Examples include tests of the distance duality relation based on comparisons between luminosity and angular-diameter distances~\cite{Hu:2026yda,  Das:2026hfp,Luo:2025vos, Zheng:2025cgq, Kanodia:2025jqh, Zhang:2025qbs, Teixeira:2025czm, Alfano:2025gie, Xu:2022zlm, Renzi:2021xii,  Euclid:2020ojp, Hogg:2020ktc, Lopez-Hernandez:2025lbj, Afroz:2025iwo} and curvature diagnostics combining distance and expansion-rate measurements~\cite{Clarkson:2007pz, Sapone:2014nna, Arjona:2021hmg, Euclid:2021frk, Dias:2024tpf, Dinda:2025hiu, Dinda:2025svh, Marra:2017pst}. Although these approaches provide powerful probes of individual assumptions, they are generally designed to test a single consistency condition at a time, see~\cite{Euclid:2021frk, Euclid:2025bxg} and references therein. As a result, when a discrepancy is observed between different datasets, it is often difficult to determine whether its origin lies in the distance ladder calibration, a violation of the distance duality relation, assumptions about spatial curvature, or the standard-ruler calibration itself. A framework capable of separating these effects within a common observational analysis would therefore provide a valuable diagnostic tool for interpreting cosmological tensions.

Baryon Acoustic Oscillation (BAO) measurements play a central role in this context. They provide geometric constraints on the expansion history through the quantities $(\alpha_\perp, \alpha_\parallel)$, which encode the ratio between observed distances and those predicted in a fiducial cosmology~\cite{Weinberg:2013agg,SDSS:2005xqv,2dFGRS:2005yhx,Blake:2011en,BOSS:2012dmf,eBOSS:2015jyv,DESI:2025zgx}. These observables depend on the sound horizon at the drag epoch, $\rd$, a standard ruler set by early-Universe physics. Under standard assumptions, $\rd$ is a constant, and different determinations of distance ratios involving BAO should therefore be mutually consistent and independent of redshift. 
This property was recently exploited in~\cite{Sapone:2026iwc} to construct a calibration-free null test of flat Friedmann-Lemaître-Robertson-Walker (FLRW) geometry written entirely in terms of the BAO shift parameters, in which the ratio $\rd/\rdf$ cancels identically, with $r_d^{\rm fid}$ being the sound horizon computed for a fiducial cosmology. In the present work we extend this idea: rather than constructing a single test in which the sound-horizon ratio cancels, we combine BAO with external distance and expansion-rate information to obtain four independent reconstructions of
$\rd /\rdf$, each sensitive to a distinct subset of physical assumptions. Specifically, we combine BAO measurements with either Type Ia supernovae (SNIa), which provide luminosity distances~\cite{Riess:2021jrx}, or cosmic chronometers (CC), which directly probe the expansion rate $H(z)$~\cite{Jimenez:2001gg}. If the standard cosmological framework is correct, all such reconstructions should yield the same redshift-independent value. Any disagreement therefore indicates a breakdown in at least one of the assumptions entering the corresponding reconstruction.

This leads to four distinct estimators of the same underlying quantity, which we refer to as the \emph{cosmic tetrarchy}: a minimal and complete decomposition of distance information into four complementary channels, each depending on a different subset of physical assumptions and observational inputs. By comparing these estimators and their redshift evolution, we can diagnose the most likely origin of possible inconsistencies in a largely model-independent way. Discrepancies between SNIa- and CC-based estimators are sensitive to issues related to the distance ladder or the distance duality relation, while differences between transverse and radial modes are sensitive to assumptions about spatial curvature and the expansion history. Deviations from constancy with redshift would instead indicate a breakdown of the standard-ruler picture.

The structure of the paper is as follows. In \autoref{sec:BAO} we introduce the BAO observables and define the relevant distance ratios. In \autoref{sec:tetrarchy} we construct the four estimators of the cosmic tetrarchy and discuss their underlying assumptions. \autoref{sec:data} presents the datasets used in this analysis. In \autoref{sec:method} we describe our methodology, including both a binned approach and a Gaussian Process reconstruction. The statistical consistency between the estimators is quantified in \autoref{sec:tension}. Our results are presented in \autoref{sec:results}, and we conclude in \autoref{sec:conclusions}.

\section{BAO observables and distance ratios}\label{sec:BAO}

BAO measurements provide geometrical constraints on the expansion history of the Universe through the distortion parameters $(\alpha_\perp, \alpha_\parallel)$, which quantify the mismatch between the true cosmology and the fiducial model used in the data analysis. These parameters are defined as
\begin{eqnarray}
\alpha_\perp(z) &=& \frac{\dm(z)}{\rd}\frac{\rdf}{\dmf(z)}\,,
\label{eq:alpha_perp}\\
\alpha_\parallel(z) &=& \frac{\dh(z)}{\rd}\frac{\rdf}{\dhf(z)}\,,
\label{eq:alpha_par}
\end{eqnarray}
where $\rd$ is the comoving sound horizon at the drag epoch, $\dh(z)=c/H(z)$ is the Hubble distance, and $\dm(z)$ is the transverse comoving distance. The superscript ``${\rm fid}$'' denotes quantities evaluated in the fiducial cosmology adopted in the BAO analysis.

The transverse comoving distance encodes the effect of spatial curvature and is related to the comoving radial distance
\begin{equation}
d_{\rm c}(z)=c\int_0^z \frac{{\rm d}z'}{H(z')}\,,
\end{equation}
through
\begin{equation}
\dm(z) = 
\begin{cases} 
\frac{c}{H_0\sqrt{\Omega_k}}\sinh\!\left(\frac{\sqrt{\Omega_k}\,H_0}{c}\,d_{\rm c}(z)\right) & \Omega_k > 0\,, \\[8pt]
d_c(z) & \Omega_k = 0\,, \\[8pt]
\frac{c}{H_0\sqrt{-\Omega_k}}\sin\!\left(\frac{\sqrt{-\Omega_k}\,H_0}{c}\,d_{\rm c}(z)\right) & \Omega_k < 0\,.
\end{cases}
\end{equation}
This distinction will play an important role in the construction of the estimators introduced below, as some of them require the assumption of spatial flatness while others do not.

The quantity $\rd/\rdf$ plays a special role because BAO measurements constrain distances only through ratios involving the sound horizon. Consequently, if independent measurements of the relevant cosmological distances are available, the BAO observables can be recast as direct estimators of $\rd/\rdf$. From the definitions of $\alpha_\perp$ and $\alpha_\parallel$, one obtains
\begin{eqnarray}
R_\perp(z) &=& \frac{\rd}{\rdf} = \frac{\dm(z)}{\dmf(z)\,\alpha_\perp(z)}\,, \label{eq:Rperp}\\
R_\parallel(z) &=& \frac{\rd}{\rdf} = \frac{\dh(z)}{\dhf(z)\,\alpha_\parallel(z)}\,. \label{eq:Rpar}
\end{eqnarray}
Under the standard cosmological picture, the sound horizon $\rd$ is a redshift-independent physical scale determined by pre-recombination physics. Consequently, both $R_\perp(z)$ and $R_\parallel(z)$ are expected to be constant in redshift and to coincide with each other. Any statistically significant deviation from these conditions would indicate either a breakdown of one or more of the underlying assumptions or the presence of unaccounted systematics in the data.

In practice, BAO observations do not provide direct measurements of $\dm(z)$ and $\dh(z)$ independently of $\rd$, but rather constrain the combinations encoded in $(\alpha_\perp,\alpha_\parallel)$. Therefore, additional observational input is required to reconstruct the quantities in~\autoref{eq:Rperp} and~\autoref{eq:Rpar}. Since both $\dm(z)$ and $\dh(z)$ can be inferred using either Type Ia supernovae or cosmic chronometer information, each of the estimators above admits two independent implementations, leading naturally to the four-element structure of the cosmic tetrarchy.

In the following section, we show how combining BAO measurements with either SNIa or cosmic chronometers data allows us to build these four distinct estimators of the same underlying quantity, each relying on a different set of assumptions.

\section{Constructing the cosmic tetrarchy}\label{sec:tetrarchy}


In this section, we construct four distinct estimators of the ratio $\rd/\rdf$ by combining BAO measurements with either Type Ia supernovae (SNIa) or cosmic chronometers (CC). Each estimator relies on a different set of observational inputs and physical assumptions, allowing observed inconsistencies to be associated with specific classes of assumptions. Together, these estimators define the \emph{cosmic tetrarchy}.

\subsection{SNIa-based estimators}

Type Ia supernovae provide measurements of the luminosity distance through the distance modulus $\mu(z)=5\log_{10}(d_{\rm L}/{\rm Mpc})+25$, after calibration via the distance ladder \cite{Riess:2016jrr}. In order to relate $d_{\rm L}(z)$ to the transverse comoving distance $\dm(z)$, one must assume the validity of the distance duality relation (DDR),
\begin{equation}
    d_{\rm L}(z) = (1+z)^2 d_{\rm A}(z) = (1+z)\,\dm (z)\,,
\end{equation}
which holds in any metric theory of gravity with photon number conservation.

Under this assumption,~\autoref{eq:Rperp} can be rewritten as
\begin{equation}
    R_\perp^{\rm SN}(z) = \frac{1}{1+z}\frac{d_{\rm L}(z)}{\dmf(z)\,\alpha_\perp(z)}\,.
\end{equation}
In order to construct the radial estimator $R_\parallel$, an additional assumption is required. If we further assume spatial flatness, $\Omega_k=0$, then $\dm(z)=d_c(z)$ and the Hubble distance can be expressed in terms of $d_{\rm L}(z)$ as
\begin{equation}
    \dh(z)=\frac{c}{H(z)}=\frac{d_{\rm L}(z)}{1+z}\left[\frac{d'_{\rm L}(z)}{d_{\rm L}(z)}-\frac{1}{1+z}\right]\,.
\end{equation}
Substituting into~\autoref{eq:Rpar}, we obtain
\begin{equation}
    R_\parallel^{\rm SN}(z) = \frac{d_{\rm L}(z)}{(1+z)\dhf(z)\,\alpha_\parallel(z)}\left[\frac{d'_{\rm L}(z)}{d_{\rm L}(z)}-\frac{1}{1+z}\right]\,.
\end{equation}
Therefore, the SNIa-based estimators rely on:
\begin{itemize}
    \item $R_\perp^{\rm SN}$: assumption of DDR;
    \item $R_\parallel^{\rm SN}$: assumptions of DDR and spatial flatness.
\end{itemize}

\subsection{Cosmic chronometer-based estimators}

Cosmic chronometers provide direct measurements of the expansion rate $H(z)$ from the differential ageing of galaxies. These measurements can be used to reconstruct the Hubble distance $\dh(z)=c/H(z)$ without additional assumptions.

This allows for a direct construction of the radial estimator:
\begin{equation}
    R_\parallel^{\rm CC}(z) = \frac{c}{H(z)\,\dhf(z)\,\alpha_\parallel(z)}\,.
\end{equation}
To construct the transverse estimator $R_\perp$, one must instead reconstruct the comoving distance $d_{\rm c}(z)$ through integration of $H(z)$. This requires assuming spatial flatness, such that $\dm(z)=d_{\rm c}(z)$, yielding
\begin{equation}
    R_\perp^{\rm CC}(z) = \frac{c}{\dmf(z)\,\alpha_\perp(z)} \int_0^z \frac{{\rm d}z'}{H(z')}\,.
\end{equation}
The CC-based estimators therefore rely on:
\begin{itemize}
    \item $R_\parallel^{\rm CC}$: no additional assumptions beyond the definition of $H(z)$;
    \item $R_\perp^{\rm CC}$: assumption of spatial flatness.
\end{itemize}
It is worth noting that $R_\parallel^{\rm CC}$ occupies a special role within the tetrarchy, as it can be reconstructed directly from BAO and CC observations without invoking either the distance duality relation or spatial flatness. It therefore provides the least assumption-dependent estimate of $\rd/\rdf$ among the four channels.

\subsection{Interpretation of the tetrarchy}

The four estimators defined above provide complementary reconstructions of the same underlying quantity, $\rd/\rdf$. In the standard cosmological picture, they are expected to be mutually consistent and independent of redshift. The diagnostic power of the cosmic tetrarchy lies not only in testing this expectation, but also in identifying which assumptions are implicated when discrepancies arise.

\begin{table*}[t]
\centering
\small
\renewcommand{\arraystretch}{1.4}
\begin{tabular*}{\textwidth}{@{\extracolsep{\fill}}llll}
\hline\hline
Estimator & Observational inputs & Additional assumptions & Sensitivity \\
\hline
$R_\perp^{\rm SN}$ &
BAO + SNIa &
Distance duality relation (DDR) &
Distance ladder, photon conservation \\

$R_\parallel^{\rm SN}$ &
BAO + SNIa &
DDR + spatial flatness &
Distance ladder, photon conservation, curvature \\

$R_\parallel^{\rm CC}$ &
BAO + CC &
None &
Expansion history \\

$R_\perp^{\rm CC}$ &
BAO + CC &
Spatial flatness &
Expansion history, curvature \\
\hline\hline
\end{tabular*}
\caption{
Summary of the four estimators of the cosmic tetrarchy. Each estimator provides an independent reconstruction of the same quantity, $\rd/\rdf$, but relies on a different combination of observational inputs and physical assumptions. The third column indicates the assumptions in addition to a standard FLRW expansion history.}
\label{tab:tetrarchy_summary}
\end{table*}

Because each estimator depends on a different combination of observational inputs and assumptions, the pattern of agreement or disagreement among them provides direct information on the origin of a potential inconsistency. For example:

\begin{itemize}
    \item discrepancies between SNIa- and CC-based estimators point to effects associated with the distance ladder calibration, photon conservation, or violations of the distance duality relation;

    \item discrepancies between transverse and radial estimators indicate tensions related to spatial curvature or to assumptions entering the reconstruction of cosmological distances;

    \item deviations from constancy with redshift indicate a breakdown of the standard-ruler picture or unaccounted observational systematics. 
\end{itemize}

The asymmetry of the tetrarchy is particularly informative. Among the four estimators, $R_\parallel^{\rm CC}$ is the only one that can be reconstructed without invoking either the distance duality relation or spatial flatness, while $R_\parallel^{\rm SN}$ relies on both assumptions. The remaining two estimators each depend on only one of these ingredients. This structure enables the assumptions underlying cosmological distance measurements to be tested in a controlled and complementary manner.

The tetrarchy is complete in the sense that it exhausts all possible reconstructions of $\rd/\rdf$ obtainable from the combination of BAO measurements with either SNIa- or CC-derived distance information. The four estimators therefore constitute the minimal set of independent channels through which assumptions about the distance ladder, the distance duality relation, spatial curvature, and the standard-ruler calibration can be jointly tested. Nevertheless, this set of estimators could be extended with additional observables. One future avenue is the inclusion of bright sirens, i.e. gravitational waves event with an electromagnetic counterpart, which would provide information on the luminosity distance without the need of a calibration. Bright sirens can indeed play a significant role in the investigation of DDR violations \cite{DeLeo:2025rhy}, while disentangling possible tensions between estimators from SN calibration systematics. However, at present time only one of such events has been observed \cite{LIGOScientific:2017vwq}, and much more complete catalogues will be needed to apply the analysis method discussed in this work.

Table~\ref{tab:tetrarchy_summary} highlights the complementary nature of the four estimators, patterns of agreement or disagreement among the estimators can be used to identify which assumptions are most likely responsible for a given inconsistency.

Among the four estimators introduced in this section, 
$R_\parallel^{\rm CC}$ occupies a uniquely simple position: its 
construction requires neither the distance duality relation nor an 
assumption on spatial flatness. The radial BAO observable 
$\alpha_\parallel(z)$ encodes the Hubble distance 
$\dh(z) = c/H(z)$ up to the sound-horizon ratio $\rd/\rdf$, while 
cosmic chronometers provide $H(z)$ directly. Their combination 
therefore yields $\rd/\rdf$ without invoking any of the 
assumptions that the tetrarchy is designed to probe. We refer to 
$R_\parallel^{\rm CC}$ as the \emph{anchor channel} of the framework.

This property has direct diagnostic consequences. A statistically 
significant deviation of $R_\parallel^{\rm CC}$ from a constant 
value, or of its derivative from zero, cannot be ascribed to a 
violation of the distance duality relation, nor to spatial 
curvature. The remaining possibilities are restricted to a 
smaller set:
\begin{itemize}
    \item systematics in the cosmic chronometer determination of 
    $H(z)$;
    \item systematics in the measurement of the BAO radial shift 
    parameter $\alpha_\parallel(z)$;
    \item a breakdown of the standard-ruler interpretation of BAO 
    measurements;
    \item modifications of the late-time expansion history not 
    captured by the FLRW cosmology, propagating into 
    $R_\parallel^{\rm CC}$ through $H(z)$.
\end{itemize}
The first two categories correspond to observational systematics; 
the latter two to physical effects beyond the standard 
cosmological model. Crucially, none of them overlaps with the 
assumptions probed by the remaining three channels.

Conversely, when $R_\parallel^{\rm CC}$ is consistent with a 
redshift-independent value while one or more of the other 
estimators exhibit statistically significant tensions, the source 
of the deviation can be unambiguously attributed to the 
assumptions specific to those estimators. A discrepancy isolated 
in $R_\perp^{\rm CC}$ would point to the spatial flatness 
assumption; tensions involving $R_\perp^{\rm SN}$ or 
$R_\parallel^{\rm SN}$ would instead implicate a violation of the distance 
duality relation, or the calibration of the distance ladder. The anchor role of $R_\parallel^{\rm CC}$ 
thus enables a triangulation among the four channels: deviations 
elsewhere can be cleanly diagnosed in terms of specific 
assumption failures only insofar as the anchor itself remains 
stable.

This logic also clarifies the boundaries of what the tetrarchy 
can achieve. The framework can be used in two ways: when 
$R_\parallel^{\rm CC}$ is consistent with the standard-ruler 
expectation, the other three channels operate as 
direct tests of DDR and flatness. Whereas, a deviation in the anchor  itself instead signals that the standard-ruler picture or one of 
the input datasets requires further investigation.

\section{Datasets}\label{sec:data}


In this work we use BAO measurements from the DESI DR2 data release~\cite{DESI:2025zgx}, complemented by a low-redshift point from BOSS~\cite{BOSS:2016wmc}, whose  parameters have been re-referenced to the DESI DR2 fiducial cosmology. These datasets provide measurements of the anisotropic BAO parameters $(\alpha_\perp, \alpha_\parallel)$ at multiple redshifts, obtained from fits to the galaxy clustering signal.

The BAO observables are used directly in the form of $(\alpha_\perp, \alpha_\parallel)$, together with their covariance matrices. These quantities encode the ratios between the true distances and those of a fiducial cosmology as 
in \autoref{eq:alpha_perp} and \autoref{eq:alpha_par}, and constitute the backbone of our construction of the $R_\perp$ and $R_\parallel$ estimators.

For the luminosity distance, we use the Pantheon+ compilation of Type Ia supernovae ~\cite{Brout:2022vxf}, calibrated with the SH0ES distance ladder. The SH0ES calibration~\cite{Riess:2021jrx} fixes the absolute magnitude scale, allowing for an absolute determination of distances and therefore enabling a direct reconstruction of the ratios entering the SNIa-based estimators. This calibrated dataset provides measurements of the distance modulus $\mu(z)$ over a wide redshift range. We convert the distance modulus to luminosity distance via
\begin{equation}\label{eq:mutodl}
    d_{\rm L}(z) = 10^{(\mu(z)-25)/5}\,{\rm Mpc}\,,
\end{equation}
and propagate the full covariance matrix of the Pantheon+ sample ($\mathbf{C}_\mu$) as
\begin{equation}\label{eq:covtrans}
   \mathbf{C}_{d_{\rm L}} = \mathbf{J}^{\rm T}\,\mathbf{C}_\mu\,\mathbf{J}\,
\end{equation}
where $J$ is the Jacobian of the transformation of \autoref{eq:mutodl}.

Finally, we use a compilation of cosmic chronometer measurements, which provide direct estimates of the Hubble parameter $H(z)$ based on the differential ageing of passively evolving galaxies~\cite{Jimenez:2001gg}. We follow the approach of \cite{Moresco:2022phi}, estimating the full covariance of cosmic chronometers, thus accounting for important systematic contributions\footnote{\url{https://gitlab.com/mmoresco/CCcovariance}}.


Given that these measurements are used to reconstruct the Hubble distance $\dh(z)=c/H(z)$ and, through integration, the comoving distance $d_{\rm c}(z)$ when required, we convert the CC dataset in measurements of $1/H(z)$. We invert the values of the data points and, similarly to \autoref{eq:covtrans}, we transform the covariance matrix with the Jacobian $J=-1/H^2$.

\section{Methodology}\label{sec:method}

In order to reconstruct the four estimators of the cosmic tetrarchy, we adopt two complementary approaches. The first is a binned analysis, which relies only on local combinations of the data and therefore minimises assumptions about the functional form of the observables. The second is a Gaussian Process (GP) reconstruction, which provides a continuous non-parametric description of the observables and their derivatives.

The use of both methods serves as an internal robustness test. The binned approach might suffer from numerical derivative due to taking derivatives of noisy data, while the GP might over-smooth the features of the data in reconstructing the function. Therefore, the binned and GP methods can be seen, respectively, as more optimistic and more conservative for what concerns the significance of features in the reconstructed functions. 

In \autoref{sec:primaries} we describe the reconstruction of the primary observables entering the tetrarchy estimators, while in \autoref{sec:derived} we discuss the propagation of uncertainties to the derived quantities.

\subsection{Primary functions}\label{sec:primaries}

The tetrarchy estimators require the quantities $d_{\rm L}(z)$, $1/H(z)$, $\alpha_\perp(z)$, $\alpha_\parallel(z)$ and, in some cases, their derivatives or integrals. Since these observables are measured at different redshifts, they must first be mapped onto a common redshift grid or reconstructed as continuous functions.

\begin{figure}
    \centering
    \includegraphics[width=0.99\linewidth]{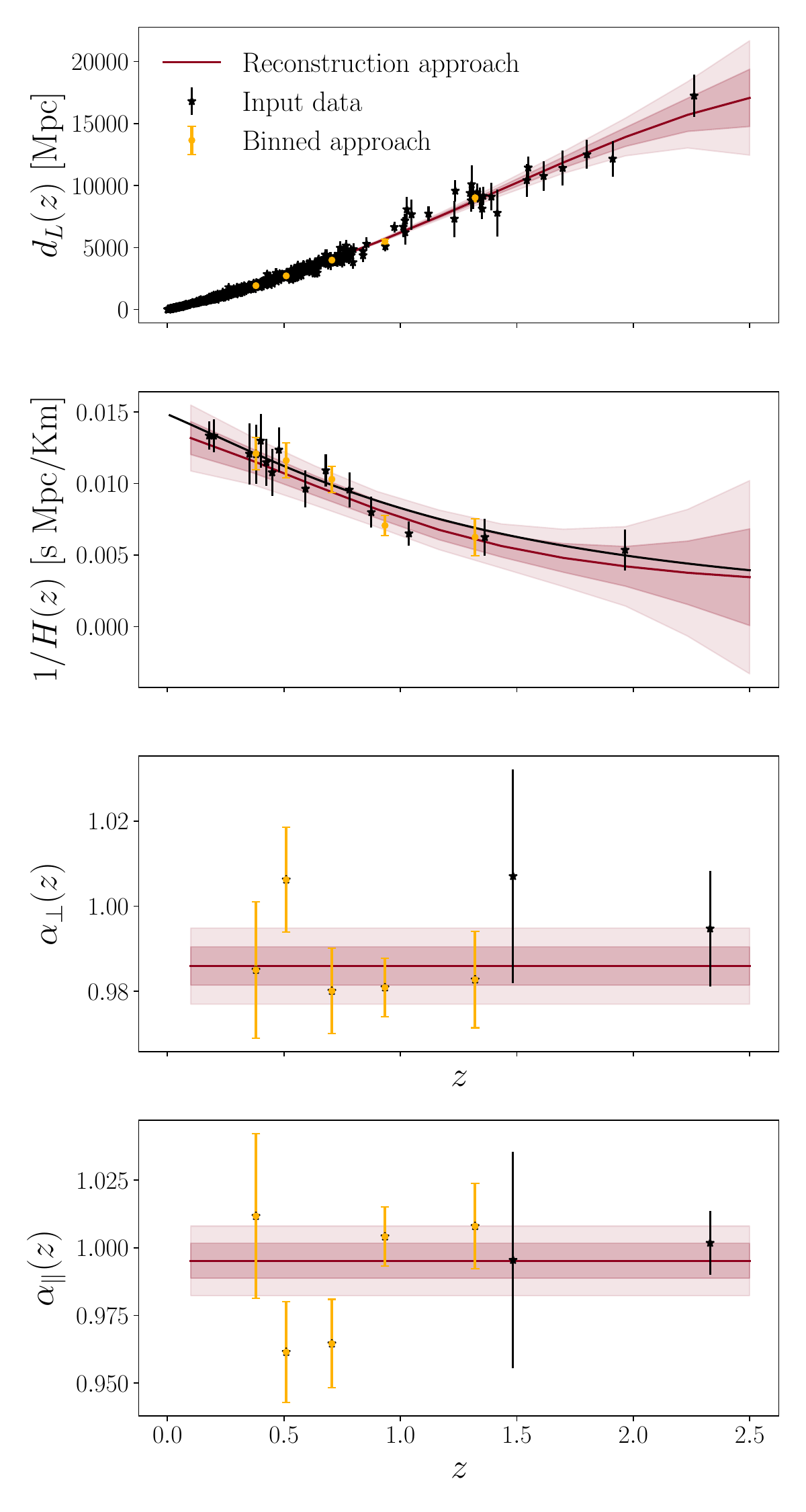}
    \caption{The yellow points and error bars show the results of the binned approach, while the red solid lines show the mean obtained with the reconstruction approach, with the shaded areas representing the 68\% and 95\% confidence regions. For the BAO observables, $\alpha_\perp$ and $\alpha_\parallel$, the binned approach coincides with the original measurements.}
    \label{fig:primaries}
\end{figure}

We therefore follow two independent approaches: a binned analysis and a Gaussian Process reconstruction. The results of both methods are shown in \autoref{fig:primaries}.

\subsubsection{Binned approach}\label{sec:binning}

In the binned approach, all observables are evaluated on the redshift grid defined by the BAO measurements.

For the SNIa sample, we compress the Pantheon+ distance moduli into redshift bins centred on the BAO redshifts using a Generalised Least Squares (GLS) estimator and the full Pantheon+ covariance matrix. For each BAO redshift $z_i$, the bin boundaries are iteratively adjusted until the GLS-weighted mean redshift of the supernovae assigned to the bin coincides with $z_i$ within a tolerance of $10^{-5}$. The full inter-bin covariance matrix is propagated throughout the analysis.

For the CC dataset, we first transform the measurements to $1/H(z)$ and assign each point to the nearest BAO redshift bin. Since CC measurements are treated as independent, the binned values are obtained through inverse-variance weighted averages.

To consistently propagate uncertainties, we assemble all primary observables into a single data vector
\begin{eqnarray}
\mathbf{x} &=& \left(\mu_1,\ldots,\mu_n,\;
                       H^{-1}_1,\ldots,H^{-1}_n,\right. \nn\\
           &&\left.
                       \alpha_{\parallel,1},\ldots,\alpha_{\parallel,n},\;
                       \alpha_{\perp,1},\ldots,\alpha_{\perp,n}
                       \right),
\end{eqnarray}
with associated covariance matrix $\mathbf{C}_{\mathbf{x}}$, assumed to be block-diagonal across datasets.

To estimate first derivatives on the irregular redshift grid, we construct a finite-difference operator matrix $\mathbf{M}$ based on second-order accurate central differences in the interior and one-sided differences at the boundaries:
\begin{eqnarray}
    \frac{df}{dz}\bigg|_{z_i}
    &\approx&
    -\frac{h_+}{h_-(h_-+h_+)}f_{i-1}
    +\frac{h_+-h_-}{h_-h_+}f_i
    \nonumber\\
    &&
    +\frac{h_-}{h_+(h_-+h_+)}f_{i+1},
\end{eqnarray}
where $h_-=z_i-z_{i-1}$ and $h_+=z_{i+1}-z_i$ account for the irregular spacing of the redshift grid.

Derivatives are obtained through matrix multiplication and their uncertainties are propagated using Jacobian matrices according to
\begin{equation}
\mathbf{C}_v
=
\mathbf{J}_v
\mathbf{C}_x
\mathbf{J}_v^{\rm T},
\end{equation}
thus preserving the full covariance information throughout the derivation chain.

For the SNIa-based estimators, we exclude the two highest BAO bins  ($z = 1.77$ and $z = 2.33$) from the binned analysis. The Pantheon+  sample provides low coverage above $z \simeq 1.5$, so the  GLS-compressed values in these bins would carry uncontrolled  uncertainties; in addition, the finite-difference operator becomes  unstable at the sparse upper boundary of the BAO grid, where the  inter-bin spacing reaches $\Delta z \simeq 0.85$. Both effects  propagate into the SNIa-based estimators, and most severely into  $R_\parallel^{\rm SN}$ through its dependence on the derivative  $d_{\rm L}'(z)$. For sake of consistency we remove these bins from all the tests.

\subsubsection{Reconstruction approach}\label{sec:reconstruction}

As a complementary approach, we reconstruct the primary functions $d_{\rm L}(z),\, 1/H(z),\, \alpha_\perp(z),\, \alpha_\parallel(z)$ using Gaussian Processes. GPs provide a non-parametric probabilistic description of a function and its covariance directly from the data.

An important property of GP reconstructions is that linear operations preserve Gaussianity. Derivatives and integrals of reconstructed functions can therefore be obtained analytically together with their associated covariance matrices. This is particularly useful for the tetrarchy estimators, which involve both derivatives of $d_{\rm L}(z)$ and integrals of $1/H(z)$.

We adopt a Radial Basis Function (RBF) kernel,
\begin{equation}
K(x,x')
=
\sigma^2
\exp\left[
-\frac{(x-x')^2}{2\ell^2}
\right],
\end{equation}
where $\sigma$ and $\ell$ are the kernel hyper-parameters. The RBF kernel is well suited to the smooth cosmological observables considered here and, being infinitely differentiable, provides stable reconstructions of the derivatives entering the tetrarchy estimators\footnote{We also tested another commonly used kernel, the Matern5/2, obtaining compatible results, except for the estimator involving the second derivative of $d_{\rm L}$. Being the Matern5/2 kernel a $C^2$ function, the second derivative exhibits numerical issues in the propagation of the uncertainty, something that is avoided when choosing the RBF which is instead $C^\infty$.}. 

The reconstruction is performed using the public code \texttt{CoRe}\footnote{\url{https://github.com/matmartinelli/CoRe}}, which implements a fully Bayesian inference of the kernel hyper-parameters and allows correlated reconstructions of multiple observables. Details of the hyper-parameter inference and the treatment of correlated functions, as well as a brief review of GP formalism, are provided in Appendix \ref{app:GPformalism}.

The reconstructed functions are evaluated at $N_{\rm recon}=10$ redshifts uniformly distributed over the interval $0.1\le z\le2.5$. We verified that increasing the reconstruction density does not affect the reconstructed means, covariances, or consistency metrics.

\subsection{Derived functions}\label{sec:derived}

Once the primary observables and their derivatives have been reconstructed, we obtain the tetrarchy estimators and their redshift derivatives through Monte Carlo propagation.

For each method, we construct a multivariate normal distribution whose mean vector is given by the reconstructed observables and whose covariance matrix is

\begin{equation}
    \mathbf{\Sigma} =
    \begin{bmatrix}
    \mathbf{C}_{d_{\rm L}} & 0 & 0 \\
    0 & \mathbf{C}_{1/H} & 0 \\
    0 & 0 & \mathbf{C}_{\alpha_\perp,\alpha_\parallel}
    \end{bmatrix}\,
\end{equation}
with $\mathbf{C}_{d_{\rm L}}$, $\mathbf{C}_{1/H}$, and $\mathbf{C}_{\alpha_\perp,\alpha_\parallel}$ being the covariance matrices of the reconstructed functions.

We then generate $N_{\rm samples}=10^5$ realisations from this distribution using \texttt{getdist}~\cite{Lewis:2019xzd}. Each realisation contains the reconstructed primary observables and all quantities required to evaluate the four tetrarchy estimators and their derivatives.

The resulting ensemble of realizations is used to estimate the means, variances, and full joint covariance matrices of all reconstructed quantities. This Monte Carlo propagation naturally preserves all correlations between the reconstructed observables. Consequently, the covariance matrices of the tetrarchy estimators contain both intra-estimator and inter-estimator correlations, which are essential for the consistency tests described in \autoref{sec:tension}.

In Appendix \ref{app:mocks}, we apply the full analysis pipeline on simulated data. This analysis is used as a validation of the approach.


\begin{figure}
    \centering
        \includegraphics[width=0.99\linewidth]{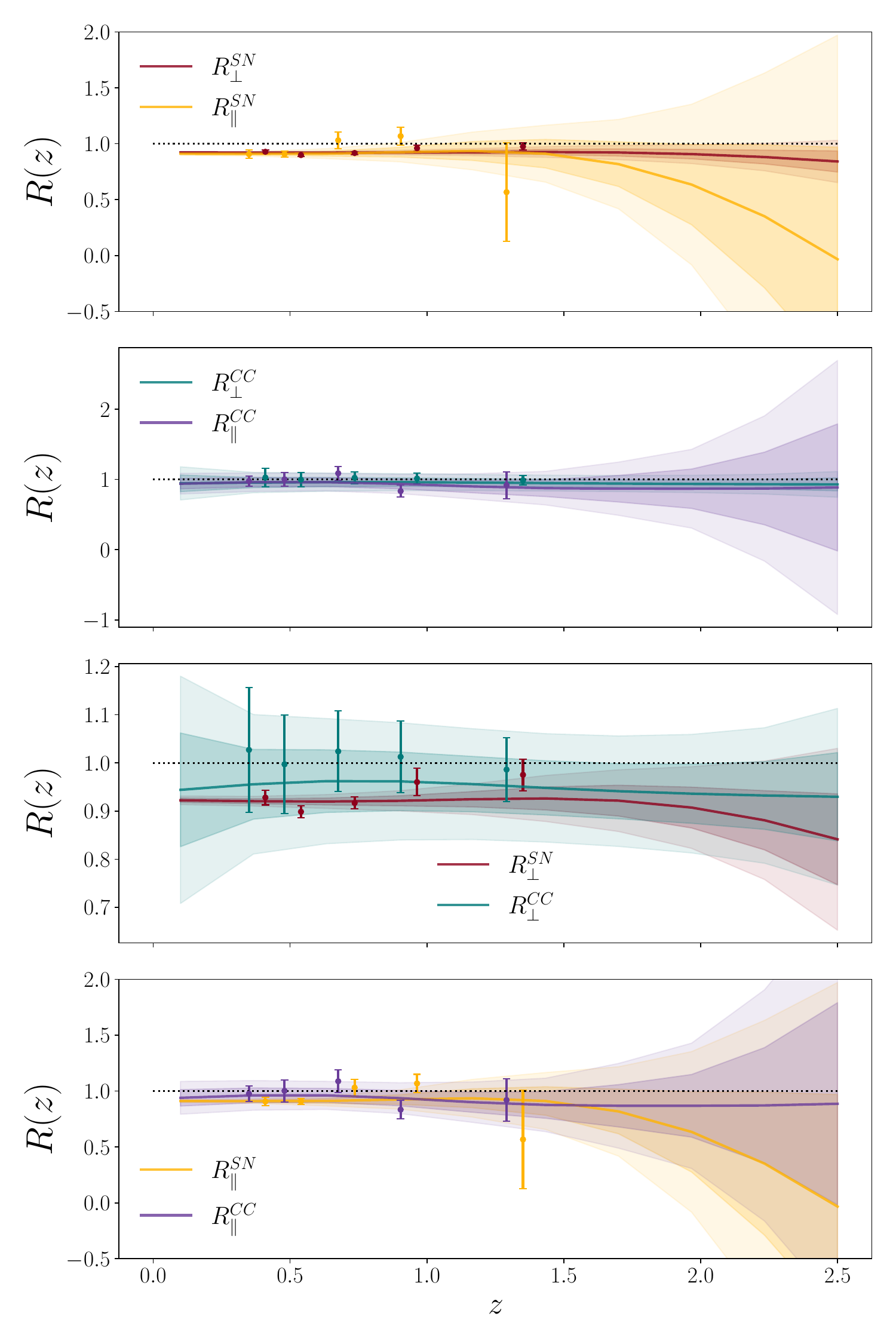}
    \caption{Reconstructed $R(z)$ functions using GP and binned approaches. The points and error bars are obtained with the latter method, while the solid lines represent the mean values of the GP with the shaded areas showing the 68\% and 95\% confidence levels.}
    \label{fig:ratios}
\end{figure}

\section{Quantification of discrepancies}\label{sec:tension}

While the reconstruction and binned approaches described in \autoref{sec:method} allows us to visualize the trends of the functions $R(z)$ and their derivatives, we need to quantify the discrepancies between the results and how much they deviate from a constant ($R'=0$).

In this section, we present the methodology used to quantitatively answer to two different questions:
\begin{itemize}
    \item Are the ratio functions in tension between them?
    \item Are the ratio compatible with the standard model expectation of $R'=0$?
\end{itemize}

To answer the first question, we compare pairs of our estimators, $R_i$ and $R_j$, reconstructed either with GP or binning at redshifts $z_{\rm recon}$, thus obtaining the vectors $\mathbf{R}_i$ and $\mathbf{R}_j$. We quantify their agreement by computing the $\chi^2$
\begin{equation}
    \chi^2_{ij}=\left[\mathbf{R}_i-\mathbf{R}_j\right]\,\mathbf{\Sigma}^{-1}_{\rm tot}\left[\mathbf{R}_i-\mathbf{R}_j\right]^{\rm T}\,,
\end{equation}
where $\mathbf{\Sigma}_{\rm tot}$ is the covariance accounting for the correlation between the two functions
\begin{equation}
    \mathbf{\Sigma}_{\rm tot} = \mathbf{\Sigma}_{ii}+\mathbf{\Sigma}_{jj}-\mathbf{\Sigma}_{ij}-\mathbf{\Sigma}_{ji}\,.
\end{equation}
In our work, accounting for this correlation is crucial, as the different functions are always related to the BAO data and, therefore, always have a non-vanishing correlation by construction.

We can then compute a $p$-value quantifying the agreement of the two reconstructed functions as
\begin{equation}
    p_{ij}=1-{\rm CDF}(\chi^2_{ij},N_{\rm dof})\,,
\end{equation}
where ${\rm CDF}$ is the cumulative density function of the $\chi^2$ distribution and $N_{\rm dof}$ is the number of degrees of freedom, i.e. the length of the reconstruction vector.

One can then define a threshold for this $p$ value, below which the two reconstructions are seen to be in tension with each other. Typically, this threshold is taken as $p_{\rm tension}=0.05$.

To answer the second question, we can use exactly the same approach and look at the reconstructed derivative of the $f$ function, comparing it with the expected vanishing value, thus
\begin{equation}
    \chi^2_{\rm const}=\mathbf{R'}\,\mathbf{\Sigma}^{-1}_{R'R'}\mathbf{R'}^{\rm T}\,,
\end{equation}
and follow the same approach as before to obtain the $p$-value assessing the compatibility with a constant function.

\section{Results}\label{sec:results}

Figure~\ref{fig:ratios} shows the four reconstructions of the ratio
$R(z)=\rd/\rdf$ obtained from the cosmic tetrarchy. The
results from the binned analysis are shown as points with error bars,
while the Gaussian Process (GP) reconstructions are shown as solid
lines with their corresponding confidence regions.

Several features are immediately apparent. First, all four estimators
are broadly compatible with a constant value of $R(z)$ over the full
redshift range covered by the data. We therefore do not find evidence for
a significant redshift evolution of the sound-horizon ratio
$\rd/\rdf$.

Second, the two CC-based estimators,
$R_\perp^{\rm CC}$ and $R_\parallel^{\rm CC}$, are mutually
consistent and remain close to a constant value. In contrast, the
two SNIa-based estimators, $R_\perp^{\rm SN}$ and $R_\parallel^{\rm SN}$, show a lower overall
normalisation, although the functions look consistent within the errors. The offset between the SNIa- and CC-based channels is
visible across most of the redshift range and is substantially larger
than the difference between the transverse and radial estimators within the
same dataset.

The uncertainties increase at high redshift, particularly for
$R_\parallel^{\rm SN}$. This reflects the reduced number of
Pantheon+ supernovae above $z\sim1.5$ and the dependence of this
estimator on the derivative of the reconstructed luminosity distance.

\begin{figure}
    \centering
        \includegraphics[width=0.99\linewidth]{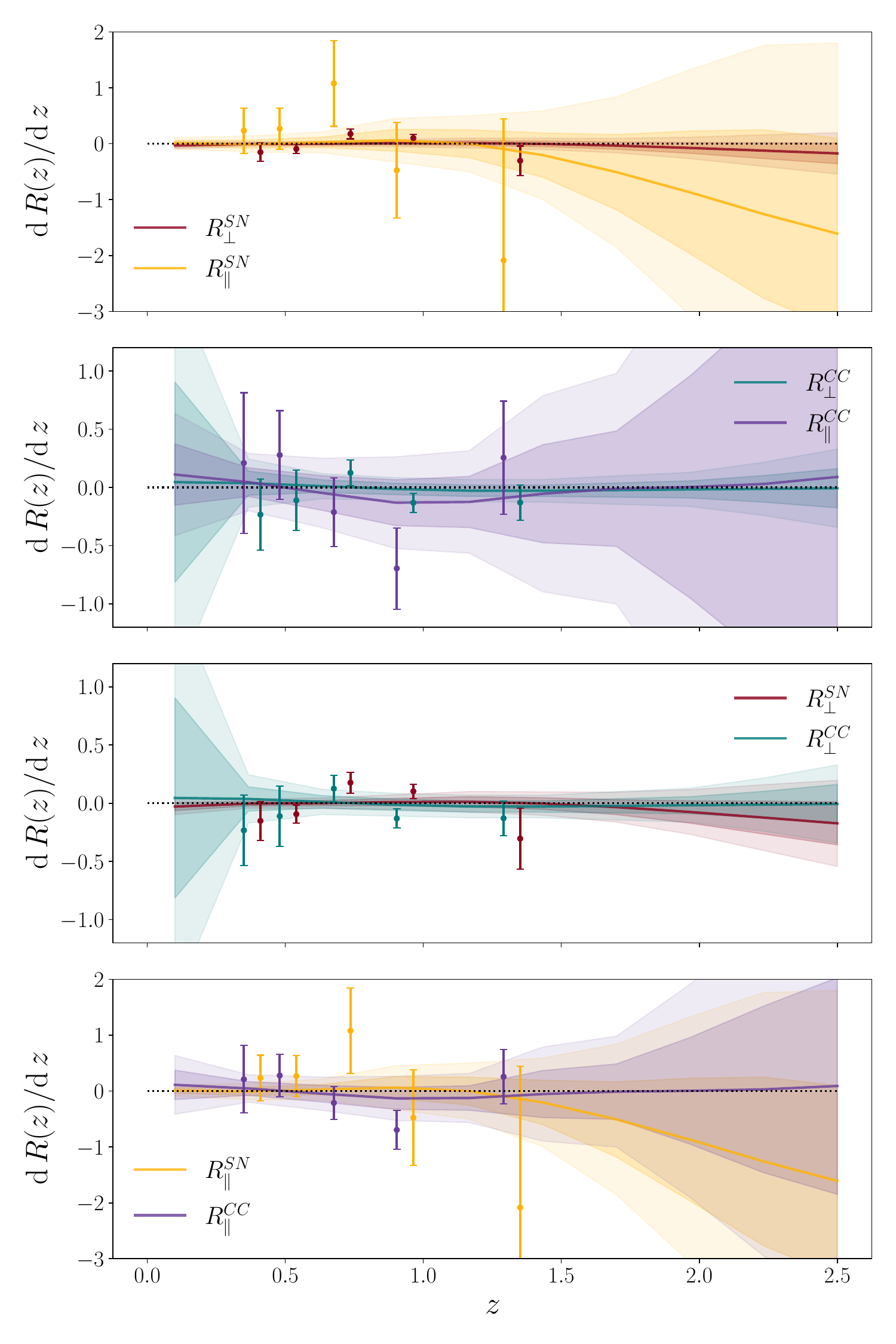}
    
    \caption{Reconstructed derivatives ${\rm d}\,R(z)/{\rm d}\,z$ using GP and binned approaches. The points and error bars are obtained with the latter method, while the solid lines represent the mean values of the GP with the shaded areas showing the 68\% and 95\% confidence levels.}
    \label{fig:derivatives}
\end{figure}

The corresponding derivatives are shown in
Fig.~\ref{fig:derivatives}. Within the uncertainties, all four
reconstructions remain consistent with the expectation
${\rm d}R/{\rm d}z = 0$. Although the uncertainties become larger at
high redshift, no estimator exhibits a statistically significant
departure from a constant value. Current data therefore provide no
evidence for a violation of the basic assumptions done in deriving the estimators, which would leave as a signature the redshift dependence of $\rd/\rdf$.

\begin{figure}
    \centering
    \includegraphics[width=0.99\linewidth]{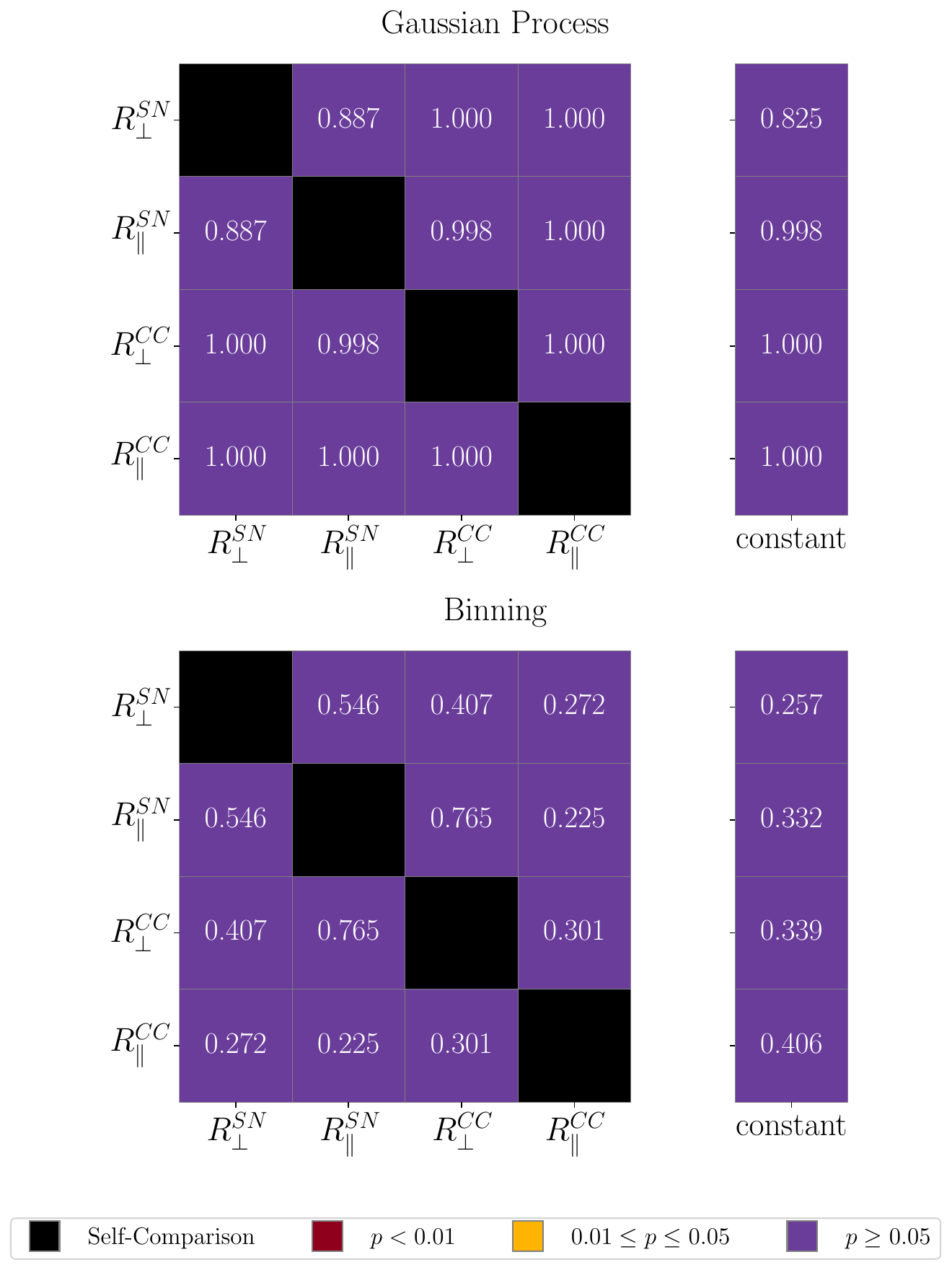}
    \caption{P-values obtained comparing the $R(z)$ functions with each other (left square matrix) and with a constant function (right column). The top panel shows the results for the GP based approach, while the bottom panel contains the results of the binned approach.}
    \label{fig:tensions_real}
\end{figure}

To quantify these visual impressions, we evaluate the pairwise
consistency of the four estimators and their compatibility with a
constant function using the methodology described in
Sec.~\ref{sec:tension}. The resulting $p$-values are reported in
Fig.~\ref{fig:tensions_real}.

For the GP reconstruction, all pairwise comparisons yield
$p$-values above the conventional $5\%$ threshold. Likewise, all
estimators are compatible with a constant value. Within the GP
framework, we therefore find no statistically significant evidence for
either internal inconsistencies among the tetrarchy estimators or a
departure from the standard-ruler expectation.

The binned analysis shows the same broad pattern, although with p-values that are systematically lower than those obtained from GP.
The comparisons
between estimators constructed from the same external dataset remain
fully consistent, both for the SNIa pair
($R_\perp^{\rm SN}$ versus $R_\parallel^{\rm SN}$) and for the CC pair
($R_\perp^{\rm CC}$ versus $R_\parallel^{\rm CC}$). The
comparisons between SNIa- and CC-based estimators yield lower
$p$-values, but in all cases this stay above the threshold of $p=0.05$.
At the same time, all four estimators remain compatible with a
constant function.

Although not statistically significant, the discrepancies arise primarily between SNIa- and
CC-based estimators, while transverse and radial estimators constructed from the same dataset exhibit higher p-values. Within the
interpretation framework developed in Sec.~\ref{sec:tetrarchy}, this pattern could point toward assumptions specific to the SNIa channels, such as the distance ladder calibration or the distance duality relation, rather than to spatial curvature.

As it can be seen in \autoref{fig:tensions_real}, there is a significant difference in the results obtained from the GP and binned approaches. This comes from how the binned and GP approaches differ in accounting for individual data points. 
The binned analysis uses a finite-difference operator that weights each measurement strongly within its own redshift bin. As a result, a single anomalous point (even one within its stated error bar) can produce an apparent local feature in the derived estimator. The propagated covariance only reflects the quoted measurement uncertainty and does not protect against this. This is again consistent with \autoref{fig:tensions_real}, where the binned analysis returns systematically lower $p$-values than the GP. The GP, by smoothing across neighbouring redshifts, is less sensitive to such single-point fluctuations, but the same smoothing also reduces sensitivity to features that may be genuinely localised in redshift.

In conclusion, these results should be treated with caution. The lower p-values in the comparison of SN and CC based estimators are worth exploring, but an improvement of the sensitivity would require a more aggressive handling of the CC systematics. Furthermore, the hints towards discrepancies appearing in the binned approach are not present in the GP analysis.
Current data do not allow us to determine whether the binned features reflect a genuine localised signal or a statistical fluctuation that the GP smoothing washes out. We therefore regard the observed SN--CC tensions as a further motivation to refine the methodology and to perform the analysis on future, more complete datasets.


\section{Conclusions}\label{sec:conclusions}

In this work, we introduced a new model-independent framework, named the \emph{cosmic tetrarchy}, to test the internal consistency of cosmological distance measurements. The method reformulates the standard cosmological framework as a single null test, namely the requirement that the dimensionless sound-horizon ratio $\rd/\rdf$ be  redshift-independent, and shows that this quantity admits four complementary reconstructions from BAO measurements combined with either Type Ia supernovae or cosmic chronometers. Because each reconstruction relies on a distinct set of physical assumptions, the pattern of agreement or disagreement among the four estimators identifies which assumption is implicated whenever an inconsistency is observed.

We applied this framework to current DESI DR2 BAO measurements, Pantheon+ supernovae calibrated by SH0ES, and cosmic chronometer compilations, performing the analysis with both a binned reconstruction and a Gaussian Process approach in order to assess the robustness of the results.

The main outcomes of the analysis, presented in Sec.~\ref{sec:results}, can be summarised at a high level as follows. 
Within current uncertainties, the cosmological distance measurements analysed here are compatible with a single, redshift-independent sound-horizon scale, and we find no statistically robust evidence for violations of the assumptions probed by the tetrarchy. The most prominent feature in the data is a systematic normalisation offset between SNIa- and CC-anchored channels, which could arise either as a manifestation of the Hubble tension or as a violation of DDR.
An additional pattern emerges in the binned reconstruction, in which 
lower p-values localise along the SN--CC axis rather than the transverse--radial one; interpreted at face value, this localisation implicates the distance duality relation or the calibration of the SNIa distance ladder rather than curvature, but the methodological caveats discussed in Sec.~\ref{sec:results} and the low statistical significance preclude a stronger conclusion with current data.

Beyond the specific datasets analysed here, the cosmic tetrarchy provides a general framework for diagnosing the origin of cosmological tensions. Future data from Euclid~\cite{Euclid:2024yrr} and the Vera C. Rubin Observatory~\cite{LSSTDarkEnergyScience:2018jkl} will substantially improve the precision of both distance and expansion-rate measurements, sharpening the sensitivity of the tests presented here. The framework also admits natural extensions to other distance probes that bypass current calibration assumptions: uncalibrated luminosity distances from gravitational-wave bright sirens, for instance, would provide an additional SNIa-independent transverse channel, while reverberation-mapped quasars offer complementary high-redshift information. Together, these directions outline a model-independent roadmap for disentangling observational systematics from possible signatures of new physics in the coming decade of precision cosmology.

\section*{Acknowledgments}
We thank Adam Riess for useful comments in the development of this work. MM acknowledges funding by the Agenzia Spaziale Italiana (\textsc{asi}) under agreement n. 2024-10-HH.0 and support from INFN/Euclid Sezione di Roma. DS acknowledges financial support from Fondecyt Regular N.~1251339

\appendix

\section{Gaussian Process formalism}\label{app:GPformalism}

For completeness, we briefly summarize the Gaussian Process formalism adopted in this work, the approach used to infer the hyper-parameters, and the handling of correlated data.

A Gaussian Process (GP) is a probability distribution over functions. Given observations $\mathbf y$ measured at positions $\mathbf X=\{x_i\}$, the GP assumption is

\begin{equation}
\mathbf y
\sim
\mathcal N
\left(
\boldsymbol{\mu},
K(\mathbf X,\mathbf X)+\mathbf{C}
\right),
\end{equation}

where $\boldsymbol{\mu}$ is the mean function, $K$ is the kernel covariance matrix, and $C$ is the observational covariance.

The predictive distribution for the function values $\mathbf f^\ast$ evaluated at positions $\mathbf X^\ast$ is obtained from the joint Gaussian distribution

\begin{equation}
\begin{bmatrix}
\mathbf y\\
\mathbf f^\ast
\end{bmatrix}
\sim
\mathcal N
\left(
\begin{bmatrix}
\boldsymbol{\mu}\\
\boldsymbol{\mu}^\ast
\end{bmatrix},
\begin{bmatrix}
K(\mathbf X,\mathbf X)+C &
K(\mathbf X,\mathbf X^\ast)
\\
K(\mathbf X^\ast,\mathbf X) &
K(\mathbf X^\ast,\mathbf X^\ast)
\end{bmatrix}
\right).
\end{equation}

Assuming vanishing means, the predictive mean and covariance are

\begin{align}
{\rm mean}(\mathbf f^\ast)
=&
K(\mathbf X^\ast,\mathbf X)
\left[
K(\mathbf X,\mathbf X)+C
\right]^{-1}
\mathbf y,
\\
{\rm Cov}(\mathbf f^\ast)
=&
K(\mathbf X^\ast,\mathbf X^\ast)
-
K(\mathbf X^\ast,\mathbf X)
\left[
K(\mathbf X,\mathbf X)+\mathbf{C}
\right]^{-1} \nonumber\\
& K(\mathbf X,\mathbf X^\ast).
\end{align}

Since differentiation and integration are linear operators, derivatives and integrals of a GP remain Gaussian Processes. The predictive mean of the first derivative is

\begin{equation}
\mathbf f^{\prime\ast}
=
\frac{\partial
K(\mathbf X^\ast,\mathbf X)}
{\partial \mathbf X^\ast}
\left[
K(\mathbf X,\mathbf X)+C
\right]^{-1}
\mathbf y.
\end{equation}

The covariance of derivatives follows directly from derivatives of the kernel,

\begin{equation}
{\rm Cov}
\left[
f'(x),
f'(x')
\right]
=
\frac{\partial^2}
{\partial x\,\partial x'}
k(x,x'),
\end{equation}

while the covariance between a function and its derivative is

\begin{equation}
{\rm Cov}
\left[
f(x),
f'(x')
\right]
=
\frac{\partial}
{\partial x'}
k(x,x').
\end{equation}

These relations are used internally by \texttt{CoRe} to reconstruct the derivatives and integrals entering the tetrarchy estimators.

\subsection{Bayesian inference of hyper-parameters}\label{app:HPinference}

The simplest approach to decide the hyper-parameters to be used in reconstructing functions with a GP is the \emph{Maximum A Posteriori} (MAP) approach. This seeks to find the specific values of the hyper-parameters $\theta = \{\ell, \sigma\}$ that maximize the posterior probability density. 
In practice, we minimize the Log-Posterior
\begin{equation}
\mathcal{L}_{MAP}(\theta) = -\log p(y | X, \theta) - \log p(\theta)\,,
\end{equation}
where $-\log p(y | X, \theta)$ is the marginal log-likelihood, which balances the model fit against the complexity penalty (determinant of the covariance matrix), and $-\log p(\theta)$ represents the priors on our hyper-parameters, preventing them from reaching unphysical values (e.g., a lengthscale of zero or infinity).

However, this method does not account for the fact that the hyper-parameters themselves will have an uncertainty. Unlike MAP, a full Bayesian approach accounts for our lack of knowledge regarding the true hyper-parameters. Instead of picking one value, we marginalize over the entire posterior distribution $p(\theta | y, X)$.

The process we follow can be summarized as:
\begin{enumerate}
\item We use a sampler (Nested Sampling via \texttt{Nautilus}\footnote{\url{https://nautilus-sampler.readthedocs.io/en/latest/}}~\cite{nautilus}) to draw thousands of samples from the hyper-parameter posterior.
\item Each sample represents a valid GP model. Some might have long length scales (smooth curves), while others have short ones (wiggly curves).
\item The final reconstruction is the weighted average of all these possible GPs.
\end{enumerate}

The uncertainty of the final reconstruction in a Bayesian GP comes from two sources:
\begin{enumerate}
\item \emph{Process Uncertainty}: the inherent variance of the GP given a fixed $\theta$;
\item \emph{Parameter Uncertainty}: the variance caused by the fact that we are not sure what $\theta$ actually is.
\end{enumerate}

The total predictive covariance is calculated as
\begin{equation}
\mathbf{\Sigma}_{total} = \mathbb{E}_\theta[\mathbf{\Sigma}_{GP}(\theta)] + \text{Var}_\theta[\mathbf{\mu}_{GP}(\theta)]\,
\end{equation}

where $\mathbb{E}_\theta[\Sigma_{GP}(\theta)]$ is the expectation value of the GP covariance over the sampled parameter space, while $\text{Var}_\theta[\mu_{GP}(\theta)]$ is the variance of the retrieved means of the GPs.

Such an approach, naturally handles heavy-tailed distributions or multimodality in the parameter space, while also producing larger error bars because it captures the uncertainty of the hyper-parameters. In this sense, this approach is more conservative than the MAP estimation, accounting for an additional error due to the hyper-parameter estimation.

\subsection{GP on correlated functions}\label{app:GPcorr}

In this work we obtained GP reconstructions of the functions $d_{\rm L}(z)$, $H(z)$, $\alpha_\perp$, and $\alpha_\parallel$. While the measurements of the first two are not correlated with any of the other functions, the same does not hold for $\alpha_\perp$ and $\alpha_\parallel$, as these are obtained from the same BAO observations. Therefore, when dealing with these functions, we need to consider not only the correlation between measurements at different redshifts, but also between different functions.

This can be handled in the context of GP, by creating a single joint observation vector $\mathbf{y}$. In general, for two observables $f_1$ and $f_2$:

\begin{equation} 
\mathbf{y} = \begin{bmatrix} \vec{y}_{f1} \\ \vec{y}_{f2} \end{bmatrix}\,,
\end{equation}

and we can represent the noise as a block-diagonal covariance matrix

\begin{equation}
    \mathbf{C} = \begin{bmatrix} \mathbf{C}_{11} & \mathbf{C}_{12} \\ \mathbf{C}_{21} & \mathbf{C}_{22} \end{bmatrix}
\end{equation}

With these settings, \texttt{CoRe} builds a joint kernel matrix $\mathbf{K}$ that describes both the auto-correlation (how $f_1$ relates to itself) and the cross-correlation (how $f_1$ relates to $f_2$).
\begin{itemize}
    \item Auto-correlation: $\text{Cov}(f_1(x), f_1(x')) = K_{11}(x, x')$
    \item Cross-correlation: $\text{Cov}(f_1(x), f_2(x')) = K_{12}(x, x')$
\end{itemize}

This ensures that the reconstruction of one function is informed by the data of the other. This GP construction implies that the final reconstruction is characterised by a set of hyper-parameters brought by the kernels describing the auto-covariance
of each latent function, as in a standard GP of a single function, while the cross-terms do not contribute any additional parameter to the reconstruction~\cite{Perenon:2021uom}.

\begin{figure}
    \centering
    \includegraphics[width=0.99\linewidth]{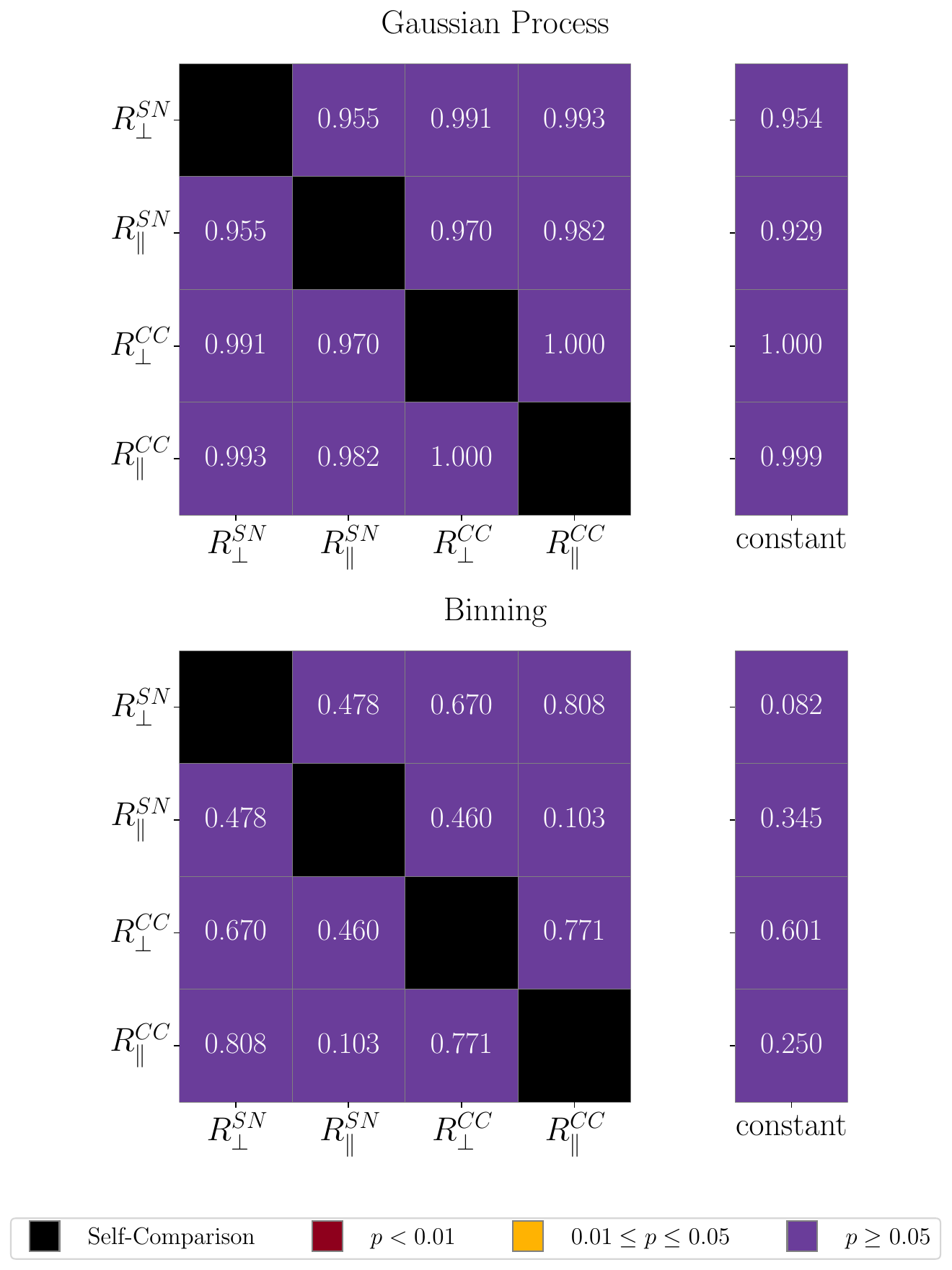}
    \caption{P-values obtained comparing the $R(z)$ functions with each other (left square matrix) and with a constant function (right column). The top panel shows the results for the GP based approach, while the bottom panel contains the results of the binned approach. These results are obtained when analyzing simulated data.}
    \label{fig:tensions_mock}
\end{figure}

\section{Synthetic data}\label{app:mocks}

In order to test our approaches, we obtain results also on a set of synthetic data. In this case, the cosmological model is known and we can check if the methods we employ return the expected results.

We build our synthetic datasets to be as close as possible to the real data we discuss in \autoref{sec:data}. In order to do so, we use the same covariances for all three datasets (SNIa, BAO, CC), and we assume the observations to be made at the same redshifts. What we change is the central value of such observations. We computed the prediction for the observed quantities ($d_{\rm L}(z)$, $H(z)$, $\alpha_\perp(z)$ and $\alpha_\parallel$) in a flat-$\Lambda$CDM fiducial model. We then use these predictions as means of a multivariate normal distribution,
extracting a realization using the covariances of the original data\footnote{We assume a diagonal covariance for the simulated cosmic chronometers dataset.}.

The dataset created with this approach share the same standard cosmology and therefore we expect to find all $R(z)$ functions to be consistent with each other, and for their derivatives to vanish.

We show the results obtained in \autoref{fig:tensions_mock}. 
The pattern of $p$-values matches the expectation of a self-consistent dataset. For both reconstruction approaches, all pairwise comparisons between estimators yield $p$-values well above the conventional $5\%$ tension threshold, and all four estimators are individually compatible with a redshift-independent value. 

These results validate the analysis pipeline: when applied to data that by construction satisfy all the tetrarchy's assumptions, both the central reconstructions and the covariance propagation return the expected null behaviour.

\printcredits

\bibliographystyle{elsarticle-num}
\bibliography{references}

\end{document}